\documentclass{article}

\usepackage{arxiv}

\usepackage[utf8]{inputenc} 
\usepackage[T1]{fontenc}    
\usepackage{hyperref}
\hypersetup{
    colorlinks=true,
    linkcolor=blue,
    citecolor=blue
}     
\usepackage{url}            
\usepackage{booktabs}       
\usepackage{amsfonts}       
\usepackage{nicefrac}       
\usepackage{microtype}      
\usepackage{graphicx}
\usepackage[numbers,sort&compress]{natbib}
\usepackage{doi}
\usepackage{amsmath}
\usepackage{amssymb}
\usepackage{float}
\usepackage[most]{tcolorbox}
\newtcolorbox{resultbox}{
  colback=black!3, colframe=black!55, boxrule=0.5pt, arc=2pt,
  left=6pt, right=6pt, top=6pt, bottom=6pt, breakable}
\DeclareMathOperator{\sign}{sign}
  
\title{Asymptotic electromagnetic transition conditions for thin layers: the arithmetic-mean field and susceptibility hybridization}

\author{ \href{https://orcid.org/0009-0006-3388-2400}{\includegraphics[scale=0.06]{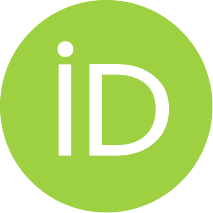}\hspace{1mm}Asbjorn M.G. Hansen} \\
	Department of Applied Mathematics\\
	University of Waterloo\\
	Waterloo, Ontario, Canada \\
	\texttt{amghanse@uwaterloo.ca}
}

\renewcommand{\headeright}{Preprint}
\renewcommand{\undertitle}{Preprint}
\renewcommand{\shorttitle}{}

\hypersetup{
pdftitle={Asymptotic electromagnetic transition conditions for thin layers: the arithmetic-mean field and susceptibility hybridization},
pdfauthor={Asbjorn M.G. Hansen},
pdfkeywords={effective transition conditions, surface susceptibility, nonlocal response, two-dimensional materials, thin-layer asymptotics},
}

\begin{document}
\maketitle

\begin{abstract}
Effective transition conditions that replace a thin polarizable layer by a zero-thickness sheet are commonly applied in electromagnetic models of two-dimensional (2D) materials and metasurfaces. These conditions usually take the arithmetic mean of the fields on either side as the polarizing field at the boundary, and characterize the layer by the $0^\mathrm{th}$ moment of its susceptibility, but both choices are typically imposed rather than derived. We derive transition conditions for a finite-thickness, in-plane homogeneous, in-plane isotropic layer with a general nonlocal susceptibility, using a single-scale asymptotic expansion in $\eta = l\max(k_0,q)$, where $l$ is the layer thickness, $k_0$ the free-space wavenumber and $q$ the in-plane wavenumber. The sheet limit must be taken after resumming the out-of-plane depolarizing field, as this ensures that the resulting sheet susceptibility is independent of the material thickness. The arithmetic mean then emerges as the leading-order polarizing field for general susceptibility profiles. Asymmetries in the layer profile modify the effective susceptibility appearing in the transition conditions, rather than altering the arithmetic-mean field prescription. If the spatial profile is not even in both out-of-plane coordinates, in-plane and out-of-plane response hybridize, turning the diagonal volumetric susceptibility into a $q$-dependent effective tensor surface susceptibility with off-diagonal components. Mirror symmetry alone does not remove this hybridization. Comparison with exact solutions for local and separable layers shows that the derived conditions remain accurate to $O(\eta)$ across propagating and evanescent excitations, whereas the standard sheet conditions fail once hybridization becomes appreciable.

\end{abstract}

\keywords{effective transition conditions, surface susceptibility, nonlocal response, two-dimensional materials, metasurfaces, generalized sheet transition conditions, spatial dispersion}

\section{Introduction}
Replacing a finite material with an effective transition (or boundary, or matching) condition (ETC) has a long history in surface optics, thin-layer modelling, and, more recently, metasurface electrodynamics. Early polarization-sheet and surface-optics formulations treated optically thin layers by assigning the response to a mathematical boundary, avoiding the need to resolve the material's microscopic structure \cite{felderhof1987electromagnetic,bedeaux2004optical,idemen1987boundary,senior1987sheet}. This zero-thickness methodology has been widely adopted in the study of atomically thin materials. Graphene, monolayer transition-metal dichalcogenides (TMDs), hexagonal boron nitride (hBN), and related systems are very often described by surface conductivity or surface susceptibility tensors rather than by volumetric dielectric functions \cite{tassin2012effective,cheng2018nonlinear,majerus2018electrodynamics,xu2021optical}. This infinitesimal-sheet philosophy has even been generalized to metasurfaces and metafilms within the theory of generalized sheet transition conditions \cite{kuester2003averaged,holloway2012overview,holloway2016homogenization}. 

To establish a link between the full finite-thickness response and an ETC, often either multi-scale \cite{lebbe2022stable,10.1098/rspa.2016.0068,holloway2016homogenization} or single-scale \cite{auvray2018asymptotic,durufle2006higher,haddar2008generalized,schmidt2010asymptotic,peron2022wentzel,peron2014equivalent,chen2017two,PhysRevB.107.085124,norgren1999effective} asymptotic expansions are applied. Multi-scale expansions are most useful for materials with important internal length scales, such as the period of scatterers in metasurfaces and metafilms. In this work, we assume in-plane homogeneity; any in-plane structure requiring homogenization is relegated to the derivation of the susceptibility, which we use as an input function in our approach. As such, a single-scale expansion is sufficient to treat the out-of-plane (OOP) response. The natural choice for this scale is the material width $l$, which is unitful, so we expand our fields in the small parameter $\eta=l \max(k_0,q)$. This definition ensures that the expansion remains well-defined in both propagating and evanescent regimes. A closely related philosophy appears in Feibelman's surface-response formulation, where microscopic field structure near an interface is not resolved explicitly, but is instead represented through low-order contributions from an expansion in the small parameter $k_z z$ where $k_z$ is the OOP wavenumber and $z$ is the OOP coordinate \cite{FEIBELMAN1982287}.

Any asymptotic procedure that replaces a finite-thickness layer with an equivalent set of transition conditions must specify how the material parameters scale as the thickness vanishes. One common choice is to hold the volumetric response fixed; in that case, the susceptibility vanishes with thickness, the leading-order conditions are those of vacuum, and material response appears only at first order, accompanied by an explicit factor of the layer thickness \cite{lebbe2022stable,10.1098/rspa.2016.0068,holloway2016homogenization}. Alternatively, one may hold the sheet response fixed, forcing the $0^\mathrm{th}$ moment of the susceptibility with respect to both OOP coordinates to be independent of material thickness, in which case the leading-order conditions describe a sheet carrying a non-trivial response \cite{nazarov2015electronic}. These are not two variants of a single limit. An atomically thin material has no canonical volumetric susceptibility, since assigning one requires an arbitrary choice of thickness. In contrast, its surface susceptibility is well-defined and derived directly from microscopic response theory. We therefore adopt the fixed-moment ($0^\mathrm{th}$ moment independent of $l$) scaling throughout.

Another choice often made in ETC models is to define the macroscopic field acting on the effective susceptibility as the arithmetic mean of the fields in the limit of approach from either side of the sheet, so that the effective field at the sheet location is formally defined in vacuum as
\begin{equation}\label{eq:46}
    \mathbf{E}(z=0)\;\;=\lim_{z_1,z_2\rightarrow0}\frac12\left[\mathbf{E}(\mathbf{r}_\|,z_1)+\mathbf{E}(\mathbf{r}_\|,z_2)\right],
\end{equation}
where $z_1\leq0$ and $z_2\geq0$. This is a common \emph{ad hoc} approximation in both metasurface and two-dimensional (2D) material models \cite{xu2021optical,dremetsika2017enhanced,majerus2018electrodynamics,cheng2018nonlinear,senior1987sheet,tassin2012effective,mansuripur2004radiation,Volokitin_2020}, but still awaits rigorous justification. The necessity of this choice follows from the infinitesimal sheet limit, and can be traced to the OOP polarization. The in-plane electric field is rendered discontinuous by in-plane gradients of the OOP polarization, while the OOP electric field itself appears to contain a singular contribution generated by that same polarization \cite{asbjorn2026anexp}. The arithmetic mean prescription resolves both issues; however, it also discards the field's interior structure, and it is not obvious \textit{a priori} that doing so is legitimate. It is nonetheless physically intuitive and has been successful in applications \cite{xu2021optical,asbjorn2026anexp,dremetsika2017enhanced,holloway2005realization}. 

While most often introduced as a choice, derivations do exist in which the arithmetic mean can be extracted rather than imposed. For instance, in Refs. \cite{krahenbuhl1993thin,igarashi1998impedance,mayergoyz1995calculation} the arithmetic mean would arise as the leading-order contribution of their boundary conditions in the limit of thin layers. The drawback of those derivations is that they use local slab models with constant material coefficients. In fact, these results are essentially identical to boundary conditions for first-order ODEs across Dirac delta singularities \cite{coutinho2009unusual}. The models take no account of the field's microscopic structure, and they provide no clear conditions defining regions of validity. To the best of the authors' knowledge, no derivation of the arithmetic-mean field prescription exists for more realistic material models. In this work, we will attempt to fill that gap.

The model considered here is deliberately general: a continuous, nonlocal, finite-thickness, polarizable thin layer, homogeneous in plane, but otherwise unrestricted in its internal structure. This includes metasurfaces, provided the discrete inclusions are first in-plane homogenized to calculate an effective susceptibility, but our main focus is on applications to atomically thin materials. We formulate ETCs for such a layer, suspended in vacuum, by truncating the single-scale asymptotic expansion at leading order. In particular, we show that the conventional arithmetic mean in Eq.~\eqref{eq:46} emerges naturally as the leading-order field prescription for the boundary constitutive relation, rather than being imposed. We also show that, remarkably, the arithmetic-mean field prescription survives for susceptibilities asymmetric with respect to both OOP coordinates. Instead, asymmetry modifies the material parameters themselves, transforming an originally diagonal volumetric susceptibility into an effective surface susceptibility tensor with off-diagonal components through hybridization between the in-plane and OOP response channels.

In Sec. \ref{sc:1}, we establish the setting and asymptotic framework. In Sec. \ref{sc:constit}, we define the constitutive relations for polarization density and establish how the susceptibility is changed in the scaled coordinate system. In Sec. \ref{sc:4}, we derive the ETCs. Finally, in Sec. \ref{sc:validation}, we numerically validate our ETCs against exact solutions of Maxwell's equations.

All work in this paper is conducted in Gaussian units.

\section{Methods}\label{sc:1}
The constitutive relation for the polarization density (per unit volume) of a non-magnetic, linear material is defined in the frequency domain by
\begin{equation}\label{eq:10}
    \mathbf{P}(\mathbf{r},\omega)=\iiint_{\mathbb{R}^3}d^3\mathbf{r}'\,\overleftrightarrow{\chi}_\mathrm{irr}(\mathbf{r},\mathbf{r}',\omega)\mathbf{E}(\mathbf{r}',\omega),
\end{equation}
where $\overleftrightarrow{\chi}_\mathrm{irr}$ is the irreducible electric field susceptibility tensor, and $\mathbf{E}$ is the total electric field; we omit frequency dependence hereafter. The notion of susceptibility reducibility comes from the Feynman-diagram tradition of quantum many-body physics. It refers to the summation over all diagrams that cannot be split into multiples of existing diagrams by cutting Coulomb propagation lines. Random phase approximation (RPA) resummation is precisely the relationship between the irreducible and the reducible susceptibilities in the specific case of noninteracting irreducible response. The irreducible function is that which responds to the total field, while the reducible one responds to the external field \cite{starke2015functional,murthy2003hamiltonian,herzog2007quantum,geraedts2012line}. Solving the electromagnetic problem self-consistently, with full implementation of boundary conditions, automatically accounts for the macroscopic subset of the Coulomb resummation.

If we assert in-plane translational invariance, $\overleftrightarrow{\chi}(\mathbf{r},\mathbf{r}')=\overleftrightarrow{\chi}(\mathbf{r}_\|-\mathbf{r}'_\|;z,z')$, we can focus our considerations on the perpendicular coordinate dependence, $z$. It is then natural to treat the problem in the in-plane Fourier domain,
\begin{equation}\label{eq:14}
    \mathbf{P}(\mathbf{q},z)=\int_{\mathbb{R}}dz'\,\overleftrightarrow{\chi}_\mathrm{irr}(\mathbf{q};z,z')\mathbf{E}(\mathbf{q},z').
\end{equation}
Hereafter, we will omit the dependence on the in-plane wavevector $\mathbf{q}$ when there is no danger of confusion. We will further only consider diagonal volumetric susceptibilities,
\begin{equation}
    \overleftrightarrow{\chi}(z,z')=\left[\begin{array}{ccc}
        \chi^\mathrm{irr}_\|(z,z') & 0 & 0\\
         0& \chi^\mathrm{irr}_\|(z,z') &0\\
         0&0&\chi^\mathrm{irr}_\perp(z,z')
    \end{array}\right],
\end{equation}
interacting with p-polarized light, where the field amplitude is given in a Cartesian coordinate system by
\begin{equation}
    \mathbf{E}(z)=\left[\begin{array}{c}
           E_\|(z)\\
         0\\
         E_\perp(z)
    \end{array}\right].
\end{equation}

For finite-thickness 2D materials, widths are usually defined by convention \cite{li2018two}. For instance, the most commonly reported width of graphene is $0.335$ nm, a number derived from the interlayer spacing in graphite crystals \cite{marinopoulos2004ab,das2011electronic}. While precise numbers will vary between materials, and even between samples of the same material due to defects and environmental conditions, it is certainly true that there will always be some width $l$ beyond which the material's charge density is negligible. We, therefore, define the material thickness as this length $l$, such that the integral bounds in Eq.~\eqref{eq:14} may be simplified as
\begin{equation}
    \int_{\mathbb{R}}dz'=\int^{\infty}_{\frac{l}{2}}dz'+\int^{\frac{l}{2}}_{-\frac{l}{2}}dz'+\int^{-\frac{l}{2}}_{-\infty}dz'\approx\int^{\frac{l}{2}}_{-\frac{l}{2}}dz',
\end{equation}
without negatively impacting the model's accuracy, and we define the coordinate origin as the geometric center of the material rather than the center of charge.

Using this material width $l$, we will introduce a scaled coordinate in the OOP direction $u=z/l$. We then define each of the electromagnetic field quantities as a function of this fast coordinate $u$,
\begin{equation}
    \mathcal{Q}_i(z)\rightarrow Q_i(u),
\end{equation}
where $\mathcal{Q}_i\in\left\{\mathbf{E},\mathbf{A},\phi\right\}$, where $\mathbf{A}$ and $\phi$ are the vector and scalar potential, respectively. With these definitions, derivatives transform as
\begin{equation}\label{eq:26}
    \frac{d}{dz}=\frac{1}{l}\frac{\partial}{\partial u}.
\end{equation}
 Finally, we will impose an asymptotic expansion on all field quantities,
\begin{equation}\label{eq:45}
    Q_i(u)=\sum^\infty_{n=0} \eta^n Q_{in}(u),
\end{equation}
where $\eta=\kappa l$ is unitless, with $\kappa=\max(k_0,q)=\sqrt{q^2+\mathrm{Re}\left(k_z\right)^2}$, where $k_0$ is the free-space wavenumber, and $q$ is the in-plane wavenumber. This choice of parameter $\kappa$ ensures that the expansion remains well-defined in the propagating and evanescent regimes. When $q\leq k_0$, $\kappa=k_0$; when $q> k_0$, $\kappa=q$.

We substitute the above expansion into the governing equations for the relevant fields, then extract the reduced governing equations by matching orders of $\eta$. The transition conditions, then, can be extracted order-by-order by evaluating the jumps in the field quantities at the endpoints of the material occupied region,
\begin{equation}
    Q_i(z=l/2)-Q_i(z=-l/2)\;=\;Q_i(u=1/2)-Q_i(u=-1/2).
\end{equation}

\section{Defining the constitutive relations}\label{sc:constit}
To determine the appropriate constitutive relations, we must clarify what field acts on the susceptibilities, and how susceptibilities are modified when converting to the scaled coordinate system.

The total electric field resulting from a continuous, three-component polarization density $\mathbf{P}$ is defined as
\begin{equation}
    \mathbf{E}(\mathbf{r})=\mathbf{E}_{ \mathrm{ext}}(\mathbf{r})+\iiint_{\mathbb{R}^3}d^3\mathbf{r}'\,\overleftrightarrow{G}(\mathbf{r}-\mathbf{r}')\mathbf{P}(\mathbf{r}'),
\end{equation}
where $\overleftrightarrow{G}$ is the electric free-space dyadic Green's function \cite{math11224691}. This relation is valid everywhere; however, if the domain includes the interior of the source, the Green's function is singular in the integration region.

As before, we will consider the problem in the Fourier domain in-plane,
\begin{equation}\label{eq:32}
    \mathbf{E}(\mathbf{q},z)=\mathbf{E}_{ \mathrm{ext}}(\mathbf{q},z)+\int_{\mathbb{R}}dz'\,\overleftrightarrow{G}(\mathbf{q},z-z')\mathbf{P}(\mathbf{q},z'),
\end{equation}
where the free-space dyadic Green's function can be defined through the scalar Green's function as
\begin{equation}\label{eq:30}
    \overleftrightarrow{G}(\mathbf{r}-\mathbf{r}')=\left[k_0^2\overleftrightarrow{I}+\nabla_\mathbf{r}\nabla_\mathbf{r}\right]g(\mathbf{r}-\mathbf{r}'),
\end{equation}
where, in the in-plane Fourier domain, the scalar Green's function becomes
\begin{equation}\label{eq:31}
    g(q,z-z')=\frac{2\pi i}{\sqrt{k_0^2-q^2}}e^{i|z-z'|\sqrt{k_0^2-q^2}},
\end{equation}
with $q=|\mathbf{q}|$.

The second-order derivative present in Eq.~\eqref{eq:30} generates a Dirac delta when acting upon the absolute value in Eq.~\eqref{eq:31}. This motivates a separation of the irregular from the regular parts of the dyadic Green's function \cite{math11224691}
\begin{equation}\label{eq:77}
    \overleftrightarrow{G}(\mathbf{q},z-z')=\overleftrightarrow{G}^{\scriptscriptstyle \mathrm{R}}(\mathbf{q},z-z')-4\pi\delta(z-z')\hat{z}\hat{z},
\end{equation}
where $\overleftrightarrow{G}^{\scriptscriptstyle \mathrm{R}}$ is purely non-singular. It is tempting to associate the second term on the right-hand side of Eq.~\eqref{eq:77} with self-interactions, the kind of contributions discarded using exclusion regions in extinction theorem or in classical derivations of the Clausius-Mossotti (or Lorentz-Lorenz) relations in the bulk of a homogeneous dielectric \cite{BornWolf:1999:Book}. Certainly, the resulting self-field contribution of the dipole coincident with the observation point should not contribute to the total field \cite{1455889,PhysRevA.97.043824}.  

Upon integration in Eq.~\eqref{eq:32}, the Dirac delta in the Green's function selects the material coordinate coincident with the observation coordinate,
\begin{align}
    \mathbf{E}(z)&=\mathbf{E}_{ \mathrm{ext}}(z)+\int_{\mathbb{R}}dz'\,\left[\overleftrightarrow{G}^{\scriptscriptstyle \mathrm{R}}(z-z')-4\pi\delta(z-z')\hat{z}\hat{z}\right]\mathbf{P}(z')\\
    &=\mathbf{E}_{ \mathrm{ext}}(z)-4\pi P_\perp(z)\hat{z}+\int_{\mathbb{R}}dz'\,\overleftrightarrow{G}^{\scriptscriptstyle \mathrm{R}}(z-z')\mathbf{P}(z').\label{eq:37}
\end{align}
Although it would appear to be singular in the infinitesimal sheet limit, the second term on the right-hand side of Eq.~\eqref{eq:37}, arising from the Dirac delta in the Green's function, is not the self-field component; it is the depolarizing field \cite{1455889}. Supporting discussion appears in Appendix \ref{sc:newapp3}.

Due to the bare, unintegrated polarization contribution in Eq.~\eqref{eq:37}, the OOP component of the constitutive relation defined in Eq.~\eqref{eq:14} is cast in implicit form,
\begin{equation}\label{eq:70}
    P_\perp(z)=\int_\mathbb{R}\chi_\perp(z,z')\left[D_\perp(z')-4\pi P_\perp(z')\right]dz',
\end{equation}
where we have defined the field $D_\perp$ for the purpose of isolating the depolarizing field by collecting the remaining terms in Eq.~\eqref{eq:37},
\begin{equation}\label{eq:57}
    D_\perp(z)\equiv\hat{z}\cdot\mathbf{E}(z)+4\pi P_\perp(z)=\hat{z}\cdot\mathbf{E}_{ \mathrm{ext}}(z)+\hat{z}\cdot\int_{\mathbb{R}}dz'\,\overleftrightarrow{G}^{\scriptscriptstyle \mathrm{R}}(z-z')\mathbf{P}(z').
\end{equation}
This quantity is the electric displacement field, but it also has a well-defined meaning in electrostatics: it is the local field in a planar geometry, obtained by subtracting the second term in Eq.~\eqref{eq:37} (i.e., the depolarizing field) from the total field. Note that $D_\perp=E_\perp$ for $|z|\geq l/2$.

The implicit nature of Eq.~\eqref{eq:70} suggests the exact resummation,
\begin{equation}\label{eq:73}
    P_\perp(z)=\left(1+4\pi\hat{\chi}^\mathrm{irr}_\perp\right)^{-1}\hat{\chi}^\mathrm{irr}_\perp[D_\perp](z),
\end{equation}
where we defined the operator
\begin{equation}
    \hat{\chi}^\mathrm{irr}_\perp[f](z)=\int_\mathbb{R}\chi^\mathrm{irr}_\perp(z,z')f(z')dz'.
\end{equation}
The operator inversion in Eq.~\eqref{eq:73} can instead be captured by an equivalent susceptibility, governed by the Dyson equation,
\begin{equation}\label{eq:72}
    \chi_\perp^\mathrm{dr}(z,z'')=\chi_\perp^\mathrm{irr}(z,z'')-4\pi\int_\mathbb{R}\chi_\perp^\mathrm{irr}(z,z')\chi_\perp^\mathrm{dr}(z',z'')dz',
\end{equation}
such that the OOP polarization density takes the form,
\begin{equation}
    P_\perp(z)=\int_\mathbb{R}\chi_\perp^\mathrm{dr}(z,z')D_\perp(z')dz',
\end{equation}
where the superscript $dr$ indicates the depolarization-resummed susceptibility. Appendix \ref{sc:newapp1} gives a derivation of this relation. For this geometry, the in-plane channel needs no resummation because the depolarizing field appears in Eq.~\eqref{eq:37} only in the OOP direction.

To apply fixed-moment scaling to the susceptibilities, we must define a scaling rule. The appropriate ansatz is $\chi\sim1/l^2$, which ensures that the $0^\mathrm{th}$ moment with respect to both OOP coordinates is independent of the material thickness in the scaled coordinates. However, we must be careful about selecting the target of this scaling rule. The infinitesimal sheet ($l\rightarrow0$) should be an effective representation of the finite one. To ensure this remains the case, the appropriate target turns out to be depolarization-resummed susceptibility, rather than the irreducible susceptibility. See Appendix \ref{sc:newapp2} for justification. 

After depolarization resummation, as defined by Eq.~\eqref{eq:72}, the polarization density can be converted to scaled coordinates,
\begin{equation}\label{eq:33}
    P_\perp(u)=\frac{1}{l}\int_\mathbb{S}\Xi_\perp^\mathrm{dr}(u,u')D_\perp(u')du',
\end{equation}
where we assumed the scaling
\begin{equation}\label{eq:56}
    \chi^\mathrm{dr}_\perp(z,z')=\frac{\Xi_\perp^\mathrm{dr}(u,u')}{l^2},
\end{equation}
and introduced the scaled coordinate domain $\mathbb{S}=\left[-a,a\right]$ as the region occupied by the material, with $a=1/2$. Note that this scaling satisfies our assertion that the $0^\mathrm{th}$ moment with respect to both OOP coordinates must remain finite in the sheet limit as $l\rightarrow0$,
\begin{equation}\label{eq:51}
    \chi^{00}_{\perp/\|}=\int_{-\infty}^\infty dz\int_{-\infty}^\infty dz'\,\chi^{\mathrm{dr}/\mathrm{irr}}_{\perp/\|}(z,z')\approx\int_{-\frac{l}{2}}^\frac{l}{2} dz\int_{-\frac{l}{2}}^\frac{l}{2} dz'\,\chi^{\mathrm{dr}/\mathrm{irr}}_{\perp/\|}(z,z')=\int_{\mathbb{S}} du\int_{\mathbb{S}} du'\,\Xi^{\mathrm{dr}/\mathrm{irr}}_{\perp/\|}(u,u'),
\end{equation}
where superscript $dr$ is attached to $\perp$, and $irr$ to $\|$. The $0^\mathrm{th}$ moment is $l$-independent in the scaled coordinates, and in fact, the $0^\mathrm{th}$ moments of $\chi$ and $\Xi$ are the same.

The constitutive relation for the in-plane polarization density requires no depolarization resummation, and so is given in the scaled coordinates by
\begin{equation}\label{eq:47}
    P_\|(u)=\frac{1}{l}\int_\mathbb{S}\Xi_\|^\mathrm{irr}(u,u')E_\|(u')du'.
\end{equation}

From now on, we will omit the superscripts on $\Xi$ and assume that all in-plane susceptibilities are irreducible, and all OOP susceptibilities are depolarization-resummed.

\section{Deriving the ETCs}\label{sc:4}

Now we construct the governing equations for fields in the region encompassing our 2D material from the Maxwell-Gauss and Maxwell-Ampere equations,
\begin{align}
    &\nabla\cdot \mathbf{E}
=-4\pi\nabla\cdot \mathbf{P},\label{eq:2}\\
    &\nabla\times\mathbf{H}=-ik_0\mathbf{E}-4\pi ik_0\mathbf{P},\label{eq:1}
\end{align}
where we have used the frequency convention $e^{-i\omega t}$. We will define the electric and magnetic fields in terms of the vector and scalar potentials,
\begin{align}
    &\mathbf{E}=-\nabla\phi+ ik_0A_\perp\hat{z},\label{eq:38}\\
    &\mathbf{H}=\nabla\times\left(A_\perp\hat{z}\right),\label{eq:39}
\end{align}
where we have defined the vector potential as $\mathbf{A}=A_\perp\hat{z}$. This gauge choice is sufficient for p-polarization and simplifies the governing equations. Because Eq.~\eqref{eq:2} governs the longitudinal response, we will collect the OOP polarization contribution into the OOP electric field to extract the governing equation for $D_\perp$, as defined in Eq.~\eqref{eq:57},
\begin{equation}\label{eq:105}
    \nabla_\|\cdot \mathbf{E}_\|+\frac{\partial D_\perp}{\partial z} =-4\pi\nabla_\|\cdot \mathbf{P}_\|,
\end{equation}
where $\nabla_\parallel$ is the parallel component of the gradient. Going forward, we will work in the in-plane Fourier domain.

First, we can insert Eq.~\eqref{eq:39} into Eq.~\eqref{eq:1}, where the $z$-row is the identity
\begin{equation}\label{eq:108}
    A_\perp=-i\frac{k_0}{q^2}D_\perp.
\end{equation}
Then, we use the $z$-component of Eq.~\eqref{eq:38} and the definition of $D_\perp$ in Eq.~\eqref{eq:57} to write
\begin{equation}
    E_\perp=-\frac{\partial \phi}{\partial z}+ik_0A_\perp=D_\perp-4\pi P_\perp.
\end{equation}
We then insert Eq.~\eqref{eq:108} and use the definition $E_\|=-iq\phi$ to find
\begin{equation}\label{eq:106}
    \frac{\partial  E_\|}{\partial z}=-4\pi iq P_\perp+iq\left(1-\frac{k_0^2}{q^2}\right)D_\perp.
\end{equation}
The second and final independent governing equation is given directly by Eq.~\eqref{eq:105},
\begin{equation}\label{eq:107}
\frac{\partial D_\perp}{\partial z} =-4\pi iq P_\|-iq E_\|.
\end{equation}
Eqs.~\eqref{eq:106} and \eqref{eq:107} are the exact governing equations for the electric field. We disregarded the nonzero $x$-component of Eq.~\eqref{eq:1} because that equation governs $A_\perp$, which, as we showed, is completely determined by $D_\perp$. 

With the governing equations in hand, we can convert to the scaled coordinate system defined in Sec. \ref{sc:1},
\begin{equation}\label{eq:109}
    \frac{1}{\eta}\frac{\partial  E_\|}{\partial u}=- \frac{4\pi iq}{\eta}\int_\mathbb{S}\Xi_\perp(u,u')D_\perp(u')du'+\frac{iq}{\kappa}\left(1-\frac{k_0^2}{q^2}\right)D_\perp,
\end{equation}
\begin{equation}\label{eq:110}
\frac{1}{\eta}\frac{\partial D_\perp}{\partial u} =- \frac{4\pi iq}{\eta}\int_\mathbb{S}\Xi_\|(u,u')E_\|(u')du'-\frac{iq}{\kappa} E_\|,
\end{equation}
where we divided both equations by $\kappa$ and used the constitutive relations defined in Eqs.~\eqref{eq:33} and \eqref{eq:47} to replace the polarization densities. In this form, the leading-order contributions are evident; however, it is interesting to consider that the exact equations are indistinguishable from the leading-order equations with modified susceptibilities defined as
\begin{align}
    &\Tilde{\Xi}_\perp(u,u')=\Xi_\perp(u,u')-\frac{\eta}{4\pi \kappa}\left(1-\frac{k_0^2}{q^2}\right)\delta(u-u'),\\
    &\Tilde{\Xi}_\|(u,u')=\Xi_\|(u,u')+\frac{\eta}{4\pi \kappa}\delta(u-u').
\end{align}
So, all of the following machinery we develop for the leading order fields can be just as easily used to extract the exact solution by using the substitution $\Xi\rightarrow\Tilde{\Xi}$.

Returning to the main point, we implement the expansion defined in Eq.~\eqref{eq:45} and truncate to $0^\mathrm{th}$ order. In this case, the fields, and correspondingly the transition conditions, are explicitly defined up to order $O(\eta)$. We can extract the governing equations corresponding to these leading-order fields from Eqs.~\eqref{eq:109} and \eqref{eq:110} by multiplying both equations by $\eta$ to eliminate any small parameters in the denominator, and then setting $\eta=0$,
\begin{equation}\label{eq:6}
    \frac{\partial  E_\|}{\partial u}=- 4\pi iq\int_\mathbb{S}\Xi_\perp(u,u')D_\perp(u')du',\qquad \frac{\partial D_\perp}{\partial u} = -4\pi iq\int_\mathbb{S}\Xi_\|(u,u')E_\|(u')du'.
\end{equation}
This system of equations can be solved formally through direct integration. The standard choice would generally be the forward antiderivative operator, which is
\begin{equation}
    \hat{I}^\mathrm{fwd}[f](u)=\int_{-a}^u f(u') du'.
\end{equation}
However, this introduces an implicit forward bias. It does not introduce inaccuracies, but it makes the result harder to interpret. We will instead use the symmetric antiderivative,
\begin{equation}
    \hat{I}^\mathrm{sym}[f](u)=\frac{1}{2}\int_{-a}^u f(u') du'-\frac{1}{2}\int_{u}^a f(u') du'.
\end{equation}
This is pointwise equivalent to the forward case, but it eliminates any directional bias. Antiderivatives are defined up to a constant, so the difference between two prescriptions boils down to the constant anchor at the defined endpoint, and the symmetrically defined constant is simply more convenient.

In Appendix \ref{sc:A}, this symmetric antiderivative is applied directly to Eq.~\eqref{eq:6}, after which the resulting system of two equations is rearranged to decouple the field components. The result is the set of Fredholm-type integral equations for the field quantities defined in Eqs.~\eqref{eq:24} and \eqref{eq:25}.  However, the resulting relations only define the fields implicitly. Explicit definitions can be extracted, but only at the boundaries of the integration region, and for particular classes of susceptibilities. We can identify the condition defining this class of susceptibilities by evaluating the implicit equations for $E_\|$ and $D_\perp$ (\eqref{eq:24} and \eqref{eq:25}) at the boundaries of $\mathbb{S}$,
\begin{equation}\label{eq:29}
    E^{\pm}_{\|}=E_\|^\mathrm{av}\mp2\pi i qD_\perp^\mathrm{av}\int_{\mathbb{S}}du_1\int_{\mathbb{S}}du_2\,\Xi_\perp(u_1,u_2) \mp (2\pi q)^2\int_{\mathbb{S}}du_1\int_{\mathbb{S}}du_2\int_{-a}^{u_2}du_3\int_{\mathbb{S}}du_4\,\Delta\Xi_{\scriptscriptstyle \mathrm{E}} E_{\|}(u_4),
\end{equation}
\begin{equation}\label{eq:40}
    D^{\pm}_{\perp}=D_\perp^\mathrm{av}\mp2\pi i qE_\|^\mathrm{av}\int_{\mathbb{S}}du_1\int_{\mathbb{S}}du_2\,\Xi_\|(u_1,u_2) \mp(2\pi q)^2\int_{\mathbb{S}}du_1\int_{\mathbb{S}}du_2\int_{-a}^{u_2}du_3\int_{\mathbb{S}}du_4\,\Delta\Xi_{\scriptscriptstyle \mathrm{D}} D_{\perp}(u_4),
\end{equation}
where $E^{\pm}_{\|}=E_{\|}(\pm a)$, $D^{\pm}_\perp=D_\perp(\pm a)$, $E_\|^\mathrm{av}=(E^+_\|+E^-_\|)/2$, $D_\perp^\mathrm{av}=(D^+_\perp+D^-_\perp)/2$
\begin{equation}\label{eq:49}
    \Delta\Xi_{\scriptscriptstyle \mathrm{E}}=\Xi_\perp(u_1,u_2)\Xi_\|(u_3,u_4)-\Xi_\perp(u_1,-u_2)\Xi_\|(-u_3,u_4),
\end{equation}
and,
\begin{equation}\label{eq:50}
    \Delta\Xi_{\scriptscriptstyle \mathrm{D}}=\Xi_\|(u_1,u_2)\Xi_\perp(u_3,u_4)-\Xi_\|(u_1,-u_2)\Xi_\perp(-u_3,u_4).
\end{equation}
As was mentioned in Sec. \ref{sc:constit}, by our definition of the material width $l$, the field $D_\perp(\pm a)=E_\perp(\pm a)$ is the macroscopic electric field in the perpendicular direction, outside the material. 

The functions defined in Eqs.~\eqref{eq:49} and \eqref{eq:50} measure a kind of symmetry-controlled hybridization between the in-plane and OOP susceptibilities. One sufficient condition on the susceptibility functions for these to be zero is that both in-plane and OOP susceptibilities be even independently in both OOP coordinates. This is not to be confused with the mirror symmetry apparent in materials like graphene, as it is a stronger condition. In this case,  Eqs.~\eqref{eq:29} and \eqref{eq:40} become explicit relations for the fields at the boundaries of the material region, and it is easy to then formulate the ETCs as
\begin{equation}\label{eq:59}
    E_\|^+-E_\|^-=-4\pi iq\,\Xi^{00}_\perp\, \frac{E^+_\perp+E^-_\perp}{2},
\end{equation}
\begin{equation}\label{eq:60}
    E_\perp^+-E_\perp^-=-4\pi iq\,\Xi^{00}_\| \,\frac{E^+_\|+E^-_\|}{2},
\end{equation}
which constitute the standard ETCs, with surface susceptibilities defined as
\begin{equation}\label{eq:52}
    \Xi^{00}_{\perp/\|}=\int_{\mathbb{S}}du_1\int_{\mathbb{S}}du_2\,\Xi_{\perp/\|}(u_1,u_2).
\end{equation}
So long as the susceptibilities are doubly even and $\eta\ll1$, the standard infinitesimal sheet boundary conditions with the arithmetic mean of the fields $E_\perp$ and $E_\|$ inducing the polarization field are accurate to $O(\eta)$. It bears repeating that the quantity defined in Eq.~\eqref{eq:52} is equal to the $0^\mathrm{th}$ moment defined in standard coordinates, $\Xi^{00}=\chi^{00}$, as established in Eq.~\eqref{eq:51}. Note also that, with the electric-field ETCs defined, the magnetic-field condition follows immediately from Eq.~\eqref{eq:108}.

\subsection{General solution}\label{sc:GS}

When the susceptibilities are not doubly even, the Fredholm integral equations of Eqs.~\eqref{eq:24} and \eqref{eq:25} can still be solved for ETCs. Appendix \ref{sc:B} shows the process. The result is similar to the doubly even case, though with the appearance of effective susceptibilities renormalized through in-plane/OOP hybridization,
\begin{equation}\label{eq:79}
    E^+_\|-E^-_\|=-4\pi i q \left(\Xi^\mathrm{eff}_{\perp\perp}\frac{E^+_\perp+E^-_\perp}{2}+\Xi^\mathrm{eff}_{\perp\|}\frac{E^+_\|+E^-_\|}{2}\right),
\end{equation}
\begin{equation}\label{eq:80}
    E^+_\perp-E^-_\perp=-4\pi i q \left(\Xi^\mathrm{eff}_{\|\|}\frac{E^+_\|+E^-_\|}{2}+\Xi^\mathrm{eff}_{\|\perp} \frac{E^+_\perp+E^-_\perp}{2}\right),
\end{equation}
where the effective susceptibilities are defined as
\begin{equation}\label{eq:22}
    \Xi^{\mathrm{eff}}_{\perp\perp/\|\|}=\int_\mathbb{S} du_1\int_\mathbb{S}du_2\Xi_{\perp/\|}(u_1,u_2)\psi_{\perp/\|}(u_2),
\end{equation}
\begin{equation}
    \Xi^\mathrm{eff}_{\perp\|/\|\perp}=-2\pi iq\int_{\mathbb{S}}du_1\int_{\mathbb{S}}du_2\int_{\mathbb{S}}du_3\int_{\mathbb{S}}du_4\,\Xi_{\|/\perp}(u_1,u_2)\sign(u_2-u_3)\Xi_{\perp/\|}(u_3,u_4)\psi_{\perp/\|}(u_4).
\end{equation}
and the function $\psi_{\perp/\|}$ is the solution to the inhomogeneous Fredholm integral equation
\begin{equation}\label{eq:23}
    \psi_{\perp/\|}(u)+(2\pi q)^2\hat{S}\hat{\Xi}_{\|/\perp}\hat{S}\hat{\Xi}_{\perp/\|}[\psi_{\perp/\|}](u)=1,
\end{equation}
with the operators being defined by
\begin{equation}\label{eq:61}
    \hat{S}[f](u)=\int_{\mathbb{S}}du'\sign(u-u')f(u'),\qquad\hat{\Xi}_{\perp/\|}[f](u)=\int_{\mathbb{S}}du'\Xi_{\perp/\|}(u,u') f(u').
\end{equation}
Note that once the function $\psi(u)$ has been computed, it may be easier to evaluate $\Xi^\mathrm{eff}_{\perp\|/\|\perp}$ using $4\pi iq\Xi^\mathrm{eff}_{\|\perp/\perp\|}=\psi_{\perp/\|}^+-\psi_{\perp/\|}^-=2(\psi_{\perp/\|}^+-1)=2(1-\psi_{\perp/\|}^-)$, where $\psi_{\perp/\|}^{\pm}=\psi_{\perp/\|}(\pm a)$.

This solution is valid everywhere except at the point where the value $-1$ is in the spectrum of the operator $(2\pi q)^2\hat{S}\hat{\Xi}_{\|/\perp}\hat{S}\hat{\Xi}_{\perp/\|}$. However, the kernel in Eq.~\eqref{eq:23} inherits square integrability from its constituents, which is a sufficient condition for the Fredholm alternative \cite{kress2013linear}. This means that the point at which Eq.~\eqref{eq:23} has no solutions is exactly the point when the corresponding homogeneous equation has non-trivial solutions. 

Remarkably, the appropriate polarizing fields to order $O(\eta)$ remain the arithmetic means of $E_\perp$ and $E_\|$ in general, for a material in vacuum, while the susceptibility exhibits a symmetry-controlled hybridization between the in-plane and the OOP polarizations that can be expressed in terms of an effective surface susceptibility. Notably, this hybridization turns a diagonal volumetric susceptibility into a tensor sheet susceptibility with off-diagonal components. However, the off-diagonal components of the hybridized susceptibility vanish for mirror-symmetric profiles. This is obvious if we make use of the alternative definition $4\pi iq\Xi^\mathrm{eff}_{\|\perp/\perp\|}=\psi_{\perp/\|}^+-\psi_{\perp/\|}^-$. For mirror symmetric profiles, $\psi_{\perp/\|}(u)$ is even, so $\psi_{\perp/\|}^+-\psi_{\perp/\|}^-=0$. The overall hybridization remains, as we argued previously based on Eqs.~\eqref{eq:49} and \eqref{eq:50}, and is captured through the definition of the diagonal effective susceptibility components, which do not vanish unless the response functions are even independently in both coordinates.

This conclusion is especially relevant for modern van der Waals heterostructures and polar 2D materials. For example, in a Janus TMD such as MoSSe, the upper and lower chalcogen planes are atomically distinct, so the monolayer explicitly breaks OOP mirror symmetry \cite{liu2025anomalous}. Eqs.~\eqref{eq:79} and \eqref{eq:80} handle asymmetries in the underlying susceptibilities unambiguously. Layered ferroelectrics and interface-engineered polar multilayers are thicker but related structures. For example, recent work on van der Waals multilayers reports that OOP symmetry breaking can arise in otherwise symmetric heterostructures due to changes in the interlayer registry \cite{chen2025ferro}. These thicker systems could even be ideal candidates to test some of the finite-size effects that may come from including the first-order contribution to the expansion, but we leave that to future work.

Furthermore, according to Eq.~\eqref{eq:23}, even for highly asymmetric susceptibilities, hybridization is minimal at low $q$, and the susceptibility remains well approximated by the $0^\mathrm{th}$ moment. This is why the standard electromagnetic boundary conditions of Eqs.~\eqref{eq:59} and \eqref{eq:60} have historically been used without issue.

\subsection{Separability}\label{sc:SS}
We will also consider the important case of separable susceptibility, characterized by
\begin{equation}\label{eq:41}
    \Xi_{\perp/\|}(u,u')=\Xi_{\perp/\|}^{00}m_{\perp/\|}(u)n_{\perp/\|}(u'),
\end{equation}
where $\Xi_{\perp/\|}^{00}$ is the $0^\mathrm{th}$ moment in both coordinates, assuming normalization of the functions $m(u)$ and $n(u')$ according to $\int_{\mathbb{S}}m_{\perp/\|}(u)du=\int_{\mathbb{S}}n_{\perp/\|}(u)du=1$. The assumption of separability is reasonable, as it arises in \textit{ab initio} computations involving products of matrix elements of the Bloch wave functions for electronic states involved in the polarization of 2D materials. Inserting this definition into Eq.~\eqref{eq:23} gives us a Fredholm integral equation with a degenerate kernel \cite{2007introductory}. This type of equation admits exact solutions. As a result, the boundary conditions can be written in closed form as
\begin{equation}\label{eq:63}
    E^+_\|-E^-_\|=-4\pi i q\left(\frac{\Xi^{00}_\perp}{1+(2\pi q)^2\Xi^{00}_\perp\Xi^{00}_\|k_1k_2}\frac{E^+_\perp+E^-_\perp}{2}+\frac{2\pi iq\Xi^{00}_\perp\Xi^{00}_\|k_1}{1+(2\pi q)^2\Xi^{00}_\perp\Xi^{00}_\|k_1k_2}\frac{E^+_\|+E^-_\|}{2}\right),
\end{equation}
\begin{equation}\label{eq:64}
    E^+_\perp-E^-_\perp=-4\pi i q\left(\frac{\Xi^{00}_\|}{1+(2\pi q)^2\Xi^{00}_\perp\Xi^{00}_\|k_1k_2}\frac{E^+_\|+E^-_\|}{2}+\frac{2\pi iq \Xi^{00}_\perp\Xi^{00}_\|k_2}{1+(2\pi q)^2\Xi^{00}_\perp\Xi^{00}_\|k_1k_2}\frac{E^+_\perp+E^-_\perp}{2}\right),
\end{equation}
where,
\begin{equation}\label{eq:65}
    k_1=\iint_{\mathbb{S}^2}n_\perp(u_1)m_\|(u_2)\sign(u_1-u_2)du_1du_2,\;\;k_2=\iint_{\mathbb{S}^2}n_\|(u_3)m_\perp(u_4)\sign(u_3-u_4)du_3du_4.
\end{equation}
$k_1$ and $k_2$ may be interpreted as measures of the parity mismatch between the two functions. If both the observation and material functions are the same, $m(u)=n(u)$, then $k_1=\mathcal{A}=-k_2$, where
\begin{equation}\label{eq:66}
    \mathcal{A}=\iint_{\mathbb{S}^2}n_\perp(u_1)n_\|(u_2)\sign(u_1-u_2)du_1du_2.
\end{equation}

The assumption of separability softens the sufficient condition for making the right-hand sides in Eqs.~\eqref{eq:49} and \eqref{eq:50} vanish, and eliminating hybridization. Mirror symmetry, $m_{\perp/\|}(u)n_{\perp/\|}(u')=m_{\perp/\|}(-u)n_{\perp/\|}(-u')$, is sufficient, rather than evenness in both coordinates independently. The OOP mirror symmetry guarantees that $k_1=k_2=0$. This means rank-one approximations to monolayers of OOP mirror-symmetric 2D materials exhibit no hybridization and are described by Eqs.~\eqref{eq:59} and \eqref{eq:60}. Popular examples include graphene \cite{geim2009graphene}, hBN \cite{mueller2025full}, and 2H phase transition metal dichalcogenides \cite{voiry2014covalent}.

\section{Numerical validation}\label{sc:validation}
In this section, we validate the ETCs against exact solutions of Maxwell's equations for two classes of susceptibility: separable, $\chi(z,z')=\chi^{00}n(z)n(z')$, and local, $\chi(z,z')=\chi^{00}n(z)\,\delta(z-z')$. The two classes describe physically distinct layers and differ in how much internal freedom they give the polarization density. For separable susceptibilities, the effective response functions take the closed form in Sec. \ref{sc:SS}, and the relative asymmetry of the in-plane and OOP profiles controls hybridization. For local susceptibilities, which are not even in both OOP coordinates independently, hybridization is present even when the profiles are mirror symmetric. We define the susceptibilities in both cases to have the same $0^\mathrm{th}$ moments.

We build all of the profiles from the normalized bump function defined by
\begin{equation}\label{eq:bump}
    n(u;s)=\frac{d}{du}g(u),\;\;g(u)=\frac{1}{1+e^{f(u)}},\;\;f(u)=\frac{s}{u+\frac{1}{2}}-\frac{1}{1-\left(u+\frac{1}{2}\right)},
\end{equation}
which has unit area and vanishes, together with all of its derivatives, at the endpoints of $\mathbb{S}$ \cite{dechevsky2013smooth}. This choice of profile ensures that our assumption of contributions outside $\mathbb{S}$ being negligible holds exactly, since they are zero. For $s=1$ the profile is symmetric about $u=0$, and varying the parameter $s$ skews it. Note that Eq.~\eqref{eq:bump} is defined in the scaled coordinates. In the standard coordinates, we simply insert $u=z/l$ and divide by $l$,
\begin{equation}\label{eq:bump2}
    n(z;s)=\frac{d}{dz}g(z),\;\;g(z)=\frac{1}{1+e^{f(z)}},\;\;f(z)=\frac{s}{\frac{z}{l}+\frac{1}{2}}-\frac{1}{1-\left(\frac{z}{l}+\frac{1}{2}\right)}.
\end{equation}

\begin{figure}[ht]
    \centering
    \includegraphics[width=0.7\linewidth]{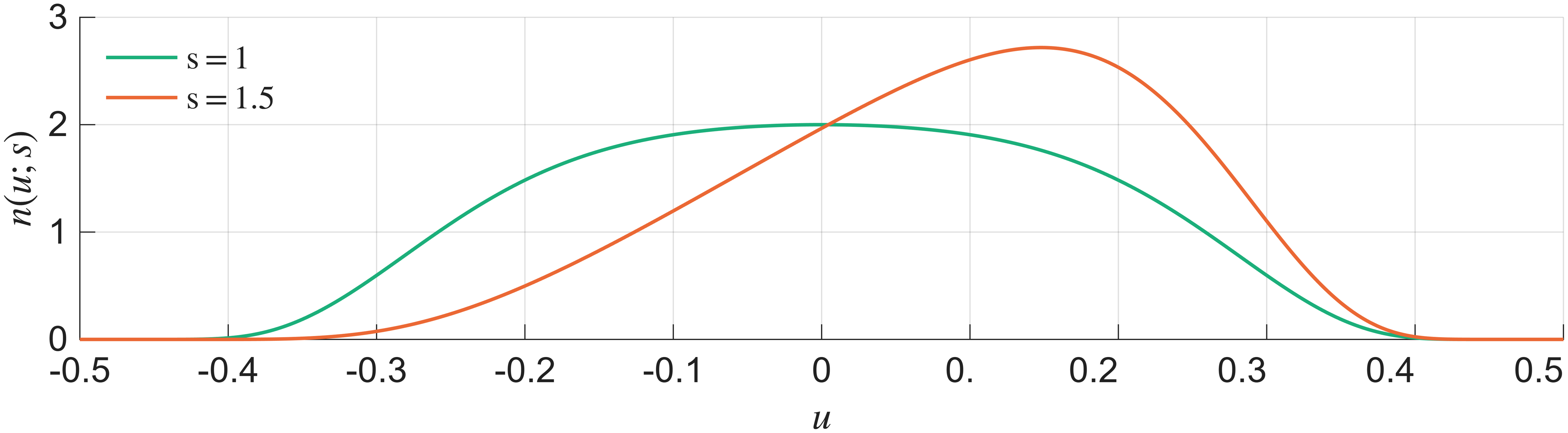}
    \caption{Eq.~\eqref{eq:bump}; a unit area bump which vanishes at the endpoints of $\mathbb{S}$ along with all derivatives.}
    \label{fig:bumps}
\end{figure}

As established in Sec. \ref{sc:constit}, we assume that $\chi_\|$ is the irreducible in-plane susceptibility and $\chi_\perp$ is the depolarization-resummed OOP susceptibility. We take $l=0.335$~nm, a free-space wavelength of $10~\mu$m ($k_0=2\pi\times10^{-4}$~nm$^{-1}$), $\chi^{00}_\|=1.56+1.16 i$~nm, and $\chi^{00}_\perp=0.15$~nm, which are the local sheet response values for undoped graphene evaluated at $\lambda=10~\mu$m~\cite{matthes2014optical}. In each case, we keep the in-plane profile centred ($s_\|=1$) and consider two OOP profiles: a mirror-symmetric profile ($s_\perp=1$) and a skewed profile ($s_\perp=1.5$). Eq.~\eqref{eq:bump} is shown in Fig. \ref{fig:bumps} for these two cases. 

We consider a p-polarized plane wave of unit amplitude incident from $z<-l/2$, and so write the exterior fields as
\begin{equation}\label{eq:valExt}
    \begin{aligned}
        E_\|&=-\frac{k_z}{k_0}\left(e^{ik_zz}-r_pe^{-ik_zz}\right), & E_\perp&=\frac{q}{k_0}\left(e^{ik_zz}+r_pe^{-ik_zz}\right), & z&<-l/2,\\
        E_\|&=-\frac{k_z}{k_0}t_pe^{ik_zz}, & E_\perp&=\frac{q}{k_0}t_pe^{ik_zz}, & z&>l/2,
    \end{aligned}
\end{equation}
with $k_z=\sqrt{k_0^2-q^2}$ and $\mathrm{Im}\,k_z\geq0$, so that $r_p$ and $t_p$ are referenced to $z=0$. We compare the models through $|r_p|^2$, using the relative error $\epsilon_R=\big||r_p|^2_\mathrm{model}-|r_p|^2_\mathrm{exact}\big|/|r_p|^2_\mathrm{exact}$, and sweep across $10^{-3}\leq\Lambda=\frac{q}{k_0}\leq10^5$.

It is important to note that while the material parameters in this section are based on graphene, the results are largely arbitrary and only serve to validate against the exact solution. We do not account for non-locality, the fact that graphene is mirror symmetric, nor do we attempt to justify our choice of bump function on physical grounds. The presence of features in the resulting plots, or the lack thereof, should not be considered representative of any particular material.

\subsection{Separable response}\label{sc:valSep}
In general, when the response is dominated by excitations that share a single vertical profile, the susceptibility is separable. We define the separable volumetric susceptibility as
\begin{equation}\label{eq:sepKernel}
    \chi_{\|/\perp}(z,z')=\chi^{00}_{\|/\perp}\,n_{\|/\perp}(z)\,n_{\|/\perp}(z'),\qquad n_{\|/\perp}(z)\equiv n(z;s_{\|/\perp}),
\end{equation}
and in the scaled coordinates as
\begin{equation}\label{eq:sepKernelScaled}
    \Xi_{\|/\perp}(u,u')=\chi^{00}_{\|/\perp}\, n_{\|/\perp}(u)\, n_{\|/\perp}(u'),\qquad  n_{\|/\perp}(u)\equiv n(u;s_{\|/\perp}),
\end{equation}
so that $\Xi^{00}_{\|/\perp}=\chi^{00}_{\|/\perp}$, consistent with Eq.~\eqref{eq:51} when one applies $\chi=\Xi/l^2$.

For a separable kernel, the shape of the polarization is fixed, and only its amplitude is unknown. We define these unknown amplitudes as
\begin{equation}\label{eq:sepP}
    P_\|(z)=\chi^{00}_\|\, n_\|(z)\,a,\qquad P_\perp(z)=\chi^{00}_\perp\, n_\perp(z)\,b,\qquad a=\int n_\|E_\|\,dz,\qquad b=\int n_\perp D_\perp\,dz.
\end{equation}
To find the exact solution, we insert these definitions into Eq.~\eqref{eq:37} and extract a system of two equations,
\begin{align}
    E_\|(z)&=E^\mathrm{ext}_\|(z)+\int\left[a\chi^{00}_\|G_{\|\|}(z,z') n_\|(z')+b\chi^{00}_\perp G_{\|\perp}(z,z') n_\perp(z')\right]dz',\label{eq:sepFields1}\\
    D_\perp(z)&=D^\mathrm{ext}_\perp(z)+\int\left[a\chi^{00}_\|G_{\perp\|}(z,z') n_\|(z')+b\chi^{00}_\perp G_{\perp\perp}(z,z') n_\perp(z')\right]dz',\label{eq:sepFields2}
\end{align}
where
\begin{equation}\label{eq:sepGreen}
    G_{\|\|}=2\pi ik_z\,e^{ik_z|z-z'|},\qquad G_{\|\perp}=G_{\perp\|}=2\pi iq\,\sign(z'-z)\,e^{ik_z|z-z'|},\qquad G_{\perp\perp}=2\pi i\frac{q^2}{k_z}\,e^{ik_z|z-z'|},
\end{equation}
are the components of the regular Green's function as introduced in Eq.~\eqref{eq:77}. The only unknown quantities here are the amplitudes $a$ and $b$. These can be solved for exactly by projecting Eqs.~\eqref{eq:sepFields1} and \eqref{eq:sepFields2} onto $ n_\|$ and $ n_\perp$, respectively, which closes the problem as a $2\times2$ linear system,
\begin{equation}\label{eq:sepSystem}
    \begin{pmatrix}1-K_{\|\|}&-K_{\|\perp}\\-K_{\perp\|}&1-K_{\perp\perp}\end{pmatrix}\begin{pmatrix}a\\ b\end{pmatrix}=\begin{pmatrix}\int n_\|E^\mathrm{ext}_\|\,dz\\ \int n_\perp D^\mathrm{ext}_\perp\,dz\end{pmatrix},\qquad K_{ij}=\chi^{00}_j\iint n_i(z)\,G_{ij}(z,z')\, n_j(z')\,dz\,dz',
\end{equation}
with $i,j\in\{\|,\perp\}$, $E^\mathrm{ext}_\|=-\frac{k_z}{k_0}e^{ik_zz}$ and $D^\mathrm{ext}_\perp=\frac{q}{k_0}e^{ik_zz}$. Evaluating Eqs.~\eqref{eq:sepFields1} at $z=-l/2$ and applying Eq.~\eqref{eq:valExt} then gives
\begin{equation}\label{eq:sepR}
    r_p=2\pi i k_0\left[a\,\chi^{00}_\|\tilde{ n}_\|+\frac{q}{k_z}b\,\chi^{00}_\perp\tilde{ n}_\perp\right],\qquad \tilde{ n}_{\|/\perp}=\int n_{\|/\perp}(z)\,e^{ik_zz}\,dz.
\end{equation}
This is the exact solution of Maxwell's equations for the separable constitutive relation, and it holds for any choice of profile.

For the ETCs, the effective response functions are given in closed form by Eqs.~\eqref{eq:63} and \eqref{eq:64}. They depend on the profiles only through $k_1$, with $k_2=-k_1$. For the mirror-symmetric layer, $k_1=0$, so $\Xi^\mathrm{eff}=\Xi^{00}$ exactly and the off-diagonal components vanish. In that case, the ETCs collapse identically to the standard conditions, Eqs.~\eqref{eq:59} and \eqref{eq:60}, for all $q$. A comparison for the symmetric profile would therefore test only the standard conditions, so we show results for the skewed profile alone. For this layer, $k_1=0.31$.

\begin{figure}[ht]
    \centering
    \includegraphics[width=0.6\linewidth]{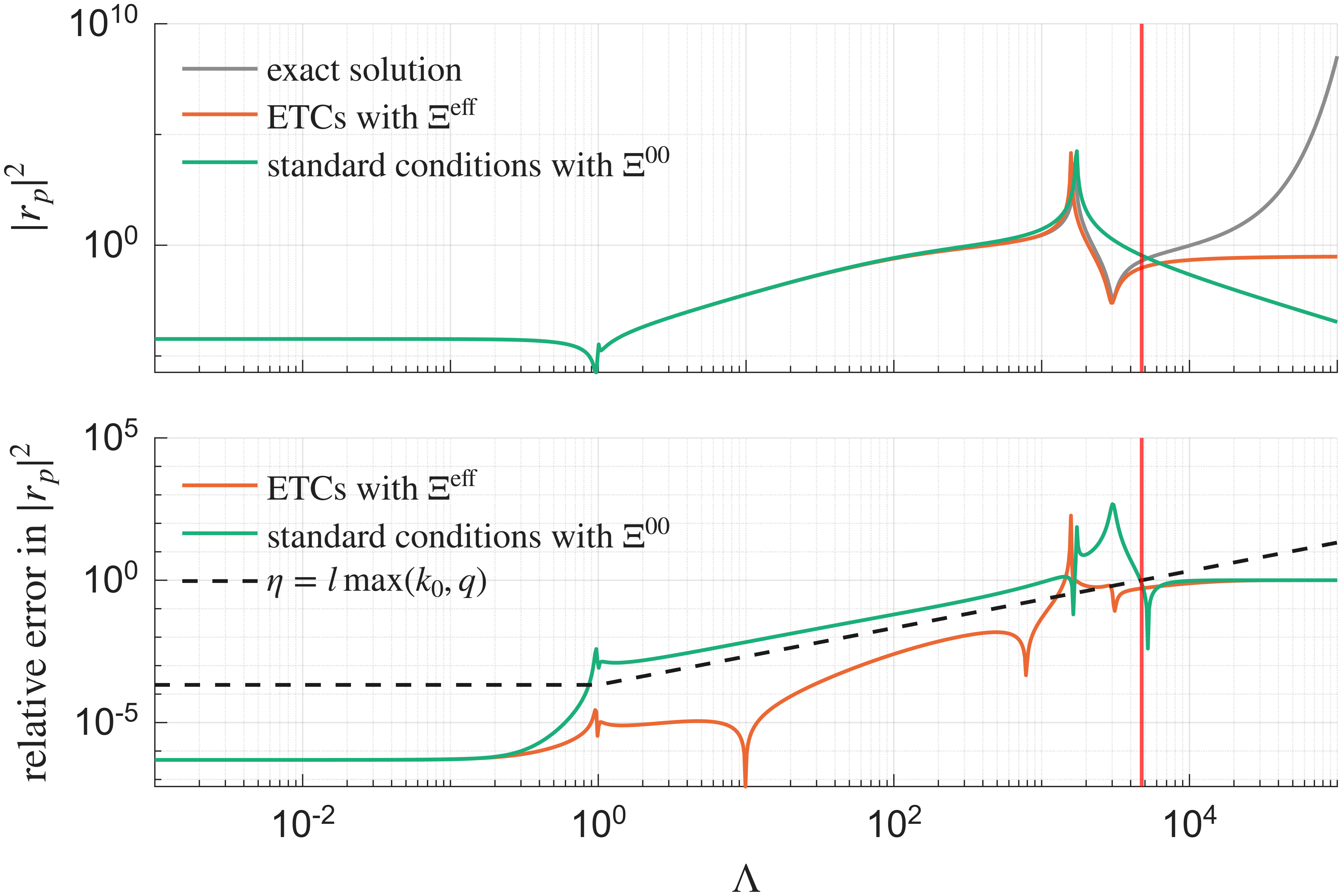}
    \caption{Separable layer with a skewed OOP profile ($s_\|=1$, $s_\perp=1.5$). (top) $|r_p|^2$ from the exact solution, the ETCs and the standard conditions. (bottom) Relative error $\epsilon_R$ in $|r_p|^2$ predicted by the ETCs, Eqs.~\eqref{eq:63} and \eqref{eq:64}, and by the standard conditions, Eqs.~\eqref{eq:59} and \eqref{eq:60}, relative to the exact solution, Eq.~\eqref{eq:sepR}. The black dashed line marks $\eta=l\max(k_0,q)$. The red vertical line indicates $\eta=1$. Parameters are given in the text.}
    \label{fig:sep}
\end{figure}

Figure~\ref{fig:sep}(top) shows that the ETCs faithfully track the exact solution right up until $\eta\approx1$. Figure~\ref{fig:sep}(bottom) shows $\epsilon_R$ over the full range of $\Lambda$. For $\Lambda\lesssim10^{-1}$, both sets of conditions give errors of the same size because hybridization is negligible at small $q$. As $\Lambda$ increases, the standard conditions, which omit hybridization, exceed $\eta$ from $\Lambda\approx1$, reaching $\epsilon_R\approx444$ at $\Lambda\approx4000$. The ETCs remain below $\eta$ up to $\Lambda\approx10^3$. However, the spike in ETC error present at $\Lambda\approx1500$ coincides with a peak in the exact solution for $|r_p|^2$ as shown in Figure~\ref{fig:sep}(top). This spike in error, therefore, is due to the offset in peak location, rather than a failure of the asymptotic expansion. Both errors saturate to 1 as the exact solution begins to grow after $\eta\approx1$.

\subsection{Local response}\label{sc:valLoc}
A local kernel describes a layer whose response at each point depends only on the field at that point. This holds when the responding charges are localized on scales small compared with the variation of the response across the layer, so that charges at different points respond independently.

In this case the local response function is defined in position coordinates as
\begin{equation}\label{eq:103}
    \chi_{\|/\perp}(z,z')=\chi^{00}_{\|/\perp}\, n_{\|/\perp}(z)\,\delta(z-z').
\end{equation}
In the scaled coordinate, we use $ n(z;s)= n(u;s)/l$ and $\delta(z-z')=\delta(u-u')/l$, which give
\begin{equation}\label{eq:104}
    \Xi_{\|/\perp}(u,u')=\chi^{00}_{\|/\perp}\, n_{\|/\perp}(u)\,\delta(u-u'),
\end{equation}
in which $ n_{\|/\perp}(u)$ and $\delta(u-u')$ are dimensionless, so that again $\Xi^{00}_{\|/\perp}=\chi^{00}_{\|/\perp}$ given $\chi=\Xi/l^2$. For the mirror-symmetric layer $\Xi^\mathrm{eff}_{\perp\|/\|\perp}=0$ and the off-diagonal terms in Eqs.~\eqref{eq:79} and \eqref{eq:80} vanish. Unlike in the separable case, however, hybridization still enters through the departure of the diagonal components of $\Xi^\mathrm{eff}$ from $\Xi^{00}$, so both profiles are shown. For the skewed layer, the off-diagonal terms are active.

For local susceptibilities, we can extract the exact response from Eqs.~\eqref{eq:109} and \eqref{eq:110} 
\begin{equation}\label{eq:valODE}
    \frac{\partial E_\|}{\partial z}=-4\pi iq\left[\chi^{00}_{\perp}\, n_{\perp}(z)-\frac{1}{4\pi}\left(1-\frac{k_0^2}{ q^2}\right)\right]D_\perp,\qquad
    \frac{\partial D_\perp}{\partial z}=-4\pi iq\Big(\chi^{00}_{\|}\, n_{\|}(z)+\frac{1}{4\pi}\Big)E_\|.
\end{equation}
We integrate it from the transmitted fields at $z=l/2$ to $z=-l/2$, and match to Eq.~\eqref{eq:valExt} to obtain $r_p$ and $t_p$ to a relative tolerance of $10^{-10}$.

The primary calculation in the ETC solution is computing $\psi_{\perp/\|}$ from Eq.~\eqref{eq:23}. Inserting Eqs.~\eqref{eq:103} and \eqref{eq:104} into Eq.~\eqref{eq:23} gives
\begin{equation}\label{eq:102}
    \psi_{\perp/\|}(u)+(2\pi q)^2\chi^{00}_{\|}\chi^{00}_{\perp}\int_{\mathbb{S}}du_1\int_{\mathbb{S}}du_2\sign(u-u_1) n_{\|/\perp}(u_1)\sign(u_1-u_2) n_{\perp/\|}(u_2)\psi_{\perp/\|}(u_2)=1,
\end{equation}
which can be converted into a system of two first-order differential equations and solved numerically. With $\psi_{\perp/\|}(u)$ in hand, the effective response functions are computed as
\begin{equation}\label{eq:100}
    \Xi^{\mathrm{eff}}_{\perp\perp/\|\|}=\chi^{00}_{\perp/\|}\int_\mathbb{S} du_1 n_{\perp/\|}(u_1)\psi_{\perp/\|}(u_1),
\end{equation}
\begin{equation}\label{eq:101}
    \Xi^\mathrm{eff}_{\perp\|/\|\perp}=-2\pi iq\chi^{00}_{\|}\chi^{00}_{\perp}\int_{\mathbb{S}}du_1\int_{\mathbb{S}}du_2\, n_{\perp/\|}(u_1)\sign(u_1-u_2)\, n_{\|/\perp}(u_2)\psi_{\|/\perp}(u_2).
\end{equation}

\begin{figure}[ht]
    \centering
    \includegraphics[width=0.8\linewidth]{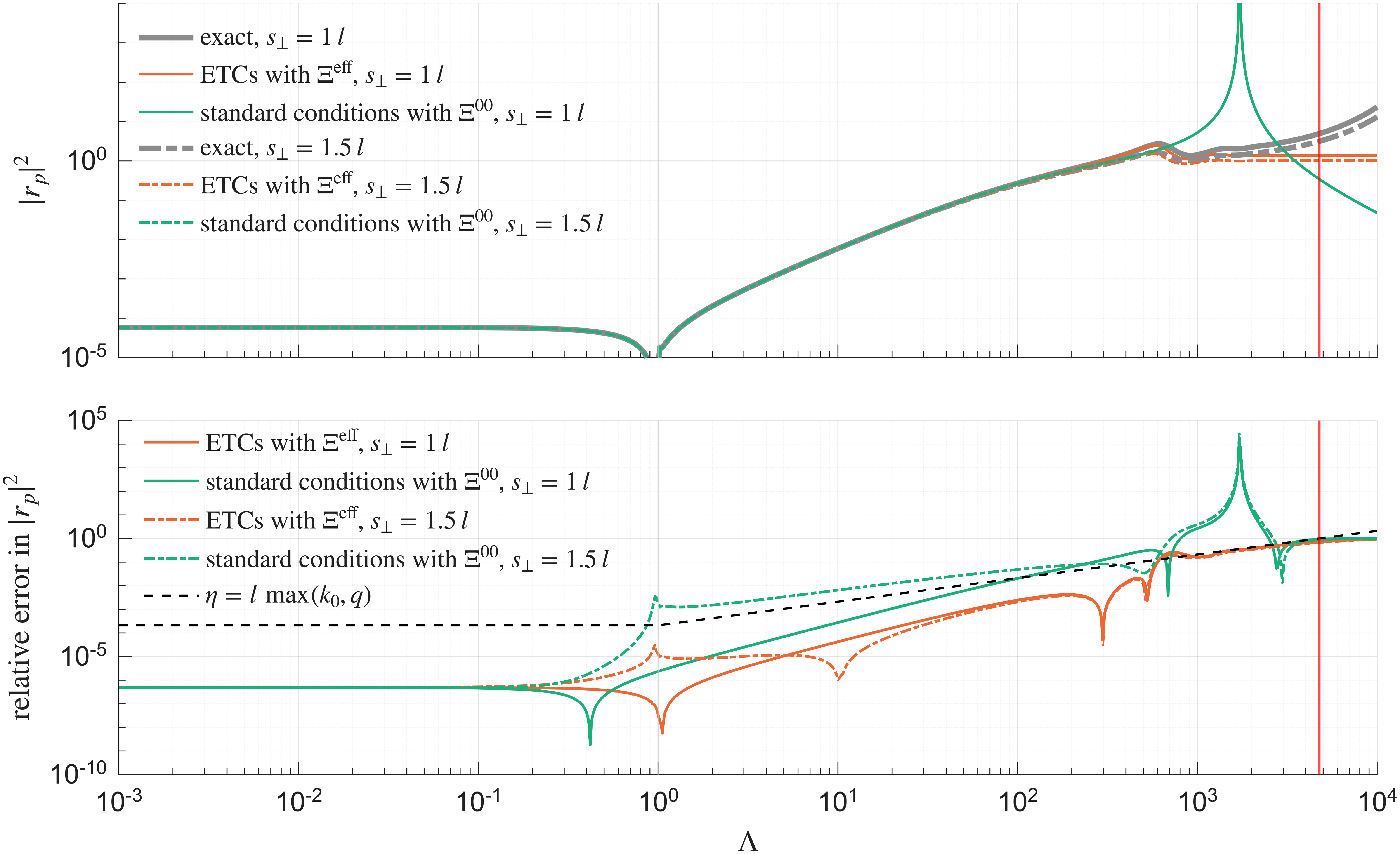}
    \caption{Local layers with a mirror-symmetric ($s_\perp=1$, solid) and a skewed ($s_\perp=1.5$, dash-dotted) OOP profile, with $s_\|=1$. (top) $|r_p|^2$ from the exact solution, the ETCs, and the standard conditions. (bottom) Relative error $\epsilon_R$ in $|r_p|^2$ predicted by the ETCs, Eqs.~\eqref{eq:79} and \eqref{eq:80}, and by the standard conditions, Eqs.~\eqref{eq:59} and \eqref{eq:60}, relative to the exact solution. The black dashed line marks $\eta=l\max(k_0,q)$. The red vertical line indicates $\eta=1$. Parameters are given in the text.} 
    \label{fig:validation}
\end{figure}

Once again, Figure~\ref{fig:validation}(top) shows a comparison between $|r_p|^2$ coming from the exact solution and the ETCs. Figure~\ref{fig:validation}(bottom) shows $\epsilon_R$ over the full range of $\Lambda$.
For $\Lambda\lesssim10^{-1}$, the ETCs and the standard conditions again give errors of the same size. As $\Lambda$ increases, $\Xi^\mathrm{eff}$ departs from $\Xi^{00}$ for both layers. In contrast to the separable case, this departure occurs for the mirror-symmetric layer as well. The standard conditions omit this hybridization, and their error exceeds $\eta$ at $\Lambda\approx100$ for the symmetric layer and $\Lambda\approx1$ for the skewed one. The ETCs remain at or below $\eta$ for the entire range of $\Lambda$ for both the symmetric layer and the skewed layer; however, both errors saturate to 1 as the exact solution begins to grow around $\eta\approx1$.

\section{Conclusion}

In this work, we derived ETCs for a finite-width thin-layer material using a single-scale asymptotic expansion. In so doing, we revealed the corresponding boundary constitutive relations for the polarization of a representative infinitesimal sheet. The infinitesimal sheet limit was defined by keeping the $0^\mathrm{th}$ moment with respect to both OOP coordinates of the susceptibility fixed, so that the finite layer approaches a nontrivial surface response as $l \rightarrow 0$. Within this scaling, and after removing the depolarizing field contribution from the polarizing field, we showed that the conventional arithmetic-mean field prescription emerges naturally as the leading-order macroscopic polarizing field, rather than as an independent modelling assumption. Our results are summarized as follows:
\begin{resultbox}
\small
\noindent\textbf{Summary of results.} To leading order in $\eta = l\,\max(k_0,q)$, the ETCs across the layer suspended in vacuum are
\begin{equation*}
E^+_\|-E^-_\|=-4\pi i q \left(\Xi^\mathrm{eff}_{\perp\perp}\frac{E^+_\perp+E^-_\perp}{2}+\Xi^\mathrm{eff}_{\perp\|}\frac{E^+_\|+E^-_\|}{2}\right),
\end{equation*}
\begin{equation*}
    E^+_\perp-E^-_\perp=-4\pi i q \left(\Xi^\mathrm{eff}_{\|\|}\frac{E^+_\|+E^-_\|}{2}+\Xi^\mathrm{eff}_{\|\perp} \frac{E^+_\perp+E^-_\perp}{2}\right),
\end{equation*}
or in terms of the magnetic field,
\begin{equation*}
  H_{\|}^{+}-H_{\|}^{-}
  = 4\pi ik_0\left(\Xi^\mathrm{eff}_{\|\|}\frac{E^+_\|+E^-_\|}{2}+\Xi^\mathrm{eff}_{\|\perp} \frac{E^+_\perp+E^-_\perp}{2}\right).
\end{equation*}
The effective susceptibilities are defined by Eq.~\eqref{eq:22}. The magnetic field condition follows immediately from the perpendicular electric field condition through Eq.~\eqref{eq:108}. When the in-plane and OOP susceptibilities are even in both OOP coordinates, the hybridization cancels, recovering the standard infinitesimal-sheet boundary conditions, Eqs.~\eqref{eq:59}--\eqref{eq:60}. When separable, the effective susceptibilities reduce to closed form, Eqs.~\eqref{eq:63}--\eqref{eq:64}.
\end{resultbox}

These results clarify that out-of-plane asymmetry does not, by itself, require replacing the arithmetic mean of the macroscopic polarizing field by some more sophisticated averaging procedure. Instead, asymmetry modifies the effective surface susceptibilities appearing in the boundary constitutive relations. The resulting framework provides a controlled way to model symmetric 2D materials, polar monolayers, asymmetric van der Waals heterostructures, and related finite-width thin-layer systems within a common effective-boundary formalism. This work considered the simplest case of leading-order contributions in vacuum surroundings. Important questions remain about how surrounding media would modify these conditions, and they will form the basis of future work focused on thin materials situated at dielectric interfaces.

Finally, applying these ETCs to real materials requires connecting the susceptibilities used here to the response functions supplied by \textit{ab initio} calculations \cite{starke2015functional,doi:10.1021/acs.nanolett.5b01251,PhysRevB.106.035431}. The susceptibilities entering the ETCs are defined with respect to the macroscopic field. \textit{Ab initio} density response functions, by contrast, typically already include some or all of the Coulomb resummation that the self-consistent Maxwell solution performs, so embedding them directly risks double counting screening. Disentangling the macroscopic screening from the microscopic local-field effects is critical to the use of \textit{ab initio} response functions in classical models. We have addressed this problem and will include it in a coming paper. Furthermore, extracting the tensor susceptibility components $\chi_\|$ and $\chi_\perp$ from a scalar density response function mixes in-plane and OOP response in a manner structurally reminiscent of the hybridization identified here. Whether these two kinds of mixing describe the same physics, or must be disentangled, will be addressed in the forthcoming work focused on \textit{ab initio} response.

\textbf{Acknowledgments}
The author wishes to express gratitude to Z. L. Mi{\v{s}}kovi{\'c}, for careful reading and thoughtful feedback.

This work was funded by the Natural Sciences and Engineering Research Council of Canada (Grant No. 2023–03397).

\section*{Declarations}
\textbf{Conflict of interest} The author declares no conflict of interest.
\bigskip

\section*{Appendices}
\appendix
\numberwithin{equation}{section}

\section{The depolarization singularity}\label{sc:newapp3}
The Dirac delta contribution to Eq.~\eqref{eq:77} is free of any $\mathbf{q}$ dependence, but the true self-interaction singularity should be a near-field effect in all three spatial coordinates; a single $z$-dependent singularity is more sheet-like than point-like. Consider the full $G_{zz}$ in the non-retarded limit ($c\rightarrow\infty$),
\begin{equation}\label{eq:76}
    G_{zz}=2\pi \left|\mathbf{q}\right|e^{-\left|\mathbf{q}\right|\left|z-z'\right|}-4\pi\delta(z-z').
\end{equation}
When $z=z'$, the first term diverges for large $\mathbf{q}$. This is exactly what we would expect to see for a divergent self-field; large $\mathbf{q}$ corresponds to increasingly short-ranged interactions. The inverse Fourier transform is conclusive. If we preselect $z=z'$ in the first term, then
\begin{equation}
    \mathcal{F}^{-1}\left\{ |\mathbf{q}| \right\}\propto\frac{1}{\rho^3},
\end{equation}
where $\mathcal{F}^{-1}$ is the inverse 2D in-plane Fourier transform, and $\rho=|\mathbf{r}_\|-\mathbf{r}'_\||$. The classic self-field divergence is actually captured by the divergence of terms linear in $|\mathbf{q}|$, rather than the Dirac delta. 

This raises the question, though: if it is not self-interaction, what does the Dirac delta singularity represent in Eq.~\eqref{eq:76}? The term itself tells us. As a direct result of the singularity in the Green's function, the total electric field in the OOP direction has a direct contribution from the bare, unintegrated polarization density given in the second term of Eq.~\eqref{eq:37}. This polarization term is precisely the source dyadic term identified in Ref. \cite{1455889}, also known as the local-field correction in electrostatics, or a generalized depolarizing field. In that paper, this depolarizing field correction takes the form $-4\pi\overleftrightarrow{L}\cdot\mathbf{P}$, where $\overleftrightarrow{L}$ is the source dyadic. This dyadic is determined by the shape of the exclusion region and compensates for the source components removed by said region in the process of handling the self-interaction singularity. In fact, $\overleftrightarrow{L}$ has been tabulated for several geometries, including pillboxes of arbitrary cross-section, for which $\overleftrightarrow{L}=\hat{z}\hat{z}$ \cite{1455889}. This exactly matches our result. We arrive at this contribution without considering any particular exclusion region because we implicitly use this kind of planar geometry when we Fourier transform in-plane. However, if one were to consider any other geometry, then the form of this contribution would change. For instance, Ref. \cite{1455889} finds $\overleftrightarrow{L}=\overleftrightarrow{I}/3$ for both cubic and spherical geometries. This would put a depolarizing field into the in-plane channel, which would force us to include an intermediate in-plane resummation as we discuss for the OOP channel in Sec. \ref{sc:constit}.

\section{Kernel resummation}\label{sc:newapp1}

There exists a unique susceptibility $\chi_\perp^\mathrm{dr}$ that captures the operator resummation and casts the polarization density in terms of the field $D_\perp$,
\begin{equation}
    P_\perp(z)=\int_\mathbb{R}\chi_\perp^\mathrm{dr}(z,z')D_\perp(z')dz'.
\end{equation}
We can insert this definition into Eq.~\eqref{eq:70},
\begin{equation}
    \int_\mathbb{R}\chi_\perp^\mathrm{dr}(z,z')D_\perp(z')dz'=\int_\mathbb{R}\chi^\mathrm{irr}_\perp(z,z')\left[D_\perp(z')-4\pi \int_\mathbb{R}\chi_\perp^\mathrm{dr}(z',z'')D_\perp(z'')dz''\right]dz',
\end{equation}
and collect factors of $D_\perp$, facilitated by a dummy variable reordering,
\begin{equation}\label{eq:71}
    \int_\mathbb{R}\left[\chi_\perp^\mathrm{dr}(z,z')-\chi^\mathrm{irr}_\perp(z,z')+4\pi\int_\mathbb{R}\chi^\mathrm{irr}_\perp(z,z'')\chi_\perp^\mathrm{dr}(z'',z')dz''\right]D_\perp(z')dz'=0.
\end{equation}
Let $R(z,z')=[...]$ be the bracketed components of the above equation. $R(z,z')$ inherits the exponential decay of each of its constituents, and so lies in $L^2$. Since Eq.~\eqref{eq:71} holds for arbitrary perturbing fields $D_\perp$, and the bracketed term is square integrable, $R(z,z')$ vanishes in $L^2$, so, by continuity of $R(z,z')$, identically \cite{2007introductory}. This gives us $R(z,z')=0$, pointwise, written in Eq.~\eqref{eq:72}.

\section{Failure of the naive sheet limit}\label{sc:newapp2}

The $1/l^2$ scaling must be applied such that the infinitesimal sheet limit ($l\rightarrow0$) remains an effective representation of the finite one. This deserves careful attention because the depolarization resummation and the infinitesimal sheet limit do not commute, and only one order corresponds to non-trivial perpendicular response. We will consider the order: sheet limit $\rightarrow$ resummation. To apply the sheet limit, let us first convert to scaled coordinates. Using the framework established in Sec. \ref{sc:1}, the irreducible OOP polarization density is transformed as
\begin{equation}
    P_\perp \left(u\right)=\frac{1}{l}\int_{\mathbb{S}}du'\,\Xi^\mathrm{irr}_\perp\left(u,u'\right)E_\perp\left(u'\right)=\frac{1}{l}\hat{\Xi}^\mathrm{irr}_\perp[E_\perp](u),
\end{equation}
where,
\begin{equation}
    \chi^\mathrm{irr}_\perp=\frac{\Xi_\perp^\mathrm{irr}}{l^2}.
\end{equation}
Inserting this transformation into Eq.~\eqref{eq:73},
\begin{align}
    P_\perp(u)&=\frac{1}{l}\left(1+\frac{4\pi}{l}\hat{\Xi}_\perp^\mathrm{irr}\right)^{-1}\hat{\Xi}^\mathrm{irr}_\perp[D_\perp](u)\\
    &=\left(l+4\pi\hat{\Xi}_\perp^\mathrm{irr}\right)^{-1}\hat{\Xi}^\mathrm{irr}_\perp[D_\perp](u).
\end{align}
This clearly corresponds to a family of different susceptibilities, rather than a rescaling of one common susceptibility. Now, if we apply the infinitesimal sheet limit, defined by $l\rightarrow0$, this reduces to
\begin{equation}
    P_\perp(u)\rightarrow\frac{1}{4\pi}\hat{\Pi}[D_\perp](u),
\end{equation}
where $\hat{\Pi}$ is the projection operator onto the range of $\hat\Xi^\mathrm{irr}_\perp$ \cite{campbell2009generalized}. The limiting value is thus independent of $\Xi_\perp^\mathrm{irr}$ regardless of the true susceptibility. Fixed-moment scaling of the irreducible kernel no longer characterizes the original material in the sheet limit. 

The consequences of improper ordering are more stark in the Dirac delta representation \cite{idemen1990universal,bedeaux2004optical,atkinson2001boundary}. In this framework, the susceptibility becomes singular in both the material and observation coordinates, $\overleftrightarrow{\chi}(z,z')=\overleftrightarrow{\chi}^{00}\delta(z)\delta(z')$ \cite{nazarov2015electronic,Despoja2017}. This then results in a singular polarization density, which we will define as
\begin{equation}
    \mathbf{P}^s(\mathbf{r})=\mathbf{P}^{\scriptscriptstyle \mathrm{2D}}(\mathbf{r}_\|)\delta(z),
\end{equation}
where $\mathbf{P}^{\scriptscriptstyle \mathrm{2D}}$ is polarization density per unit area. Inserting this polarization density into Eq.~\eqref{eq:37} gives us the total field equation in the infinitesimal sheet limit,
\begin{equation}
    \mathbf{E}(z)=\mathbf{E}_{ \mathrm{ext}}(z)-4\pi P^{\scriptscriptstyle \mathrm{2D}}_\perp\delta(z)\hat{z}+\overleftrightarrow{G}^{\scriptscriptstyle \mathrm{R}}(z)\mathbf{P}^{\scriptscriptstyle \mathrm{2D}}.
\end{equation}
As before in Eq.~\eqref{eq:57}, we can collect terms to define the total OOP electric field as 
\begin{equation}\label{eq:58}
    \hat{z}\cdot\mathbf{E}(z)=D_\perp(z)-4\pi P^{\scriptscriptstyle \mathrm{2D}}_\perp\delta(z).
\end{equation}
The terms we have encompassed in the factor $D_\perp$ are, by definition, non-singular. Inserting Eq.~\eqref{eq:58} into the OOP component of Eq.~\eqref{eq:14},
\begin{align}
    P_\perp(z)&=\int^\frac{l}{2}_{-\frac{l}{2}}dz'\,\chi^{00}_{\perp}\delta(z)\delta(z')\left[D_\perp(z')-4\pi P^{\scriptscriptstyle \mathrm{2D}}_\perp\delta(z')\right]\\
    &=\int^\frac{l}{2}_{-\frac{l}{2}}dz'\,\chi^{00}_{\perp}\left[D_\perp(z')\delta(z)\delta(z')-4\pi P^{\scriptscriptstyle \mathrm{2D}}_\perp\delta(z')\delta(z)\delta(z')\right].
\end{align}
The OOP polarization component generates undefined products of singularities, in particular, the product $\delta(z')\delta(z')$. 

Both of these examples illustrate that the infinitesimal sheet limit cannot be applied before depolarization resummation.

\section{Decoupling the leading order system}\label{sc:A}
Integrating Eq.~\eqref{eq:6} directly with the symmetric antiderivative yields,
\begin{multline}\label{eq:53}
    E_{\|}(u)=\frac{E_{\|}(a)+E_{\|}(-a)}{2}-2\pi i q\int_{-a}^udu_1\int_{\mathbb{S}}du_2\,\Xi_\perp(u_1,u_2)D_\perp(u_2)\\+2\pi i q\int^{a}_{u}du_1\int_{\mathbb{S}}du_2\,\Xi_\perp(u_1,u_2)D_\perp(u_2),
\end{multline}
\begin{multline}\label{eq:54}
    D_\perp(u)=\frac{D_\perp(a)+D_\perp(-a)}{2}-2\pi i q\int_{-a}^udu_1\int_{\mathbb{S}}du_2\,\Xi_\|(u_1,u_2)E_{\|}(u_2)\\+2\pi i q\int^{a}_{u}du_1\int_{\mathbb{S}}du_2\,\Xi_\|(u_1,u_2)E_{\|}(u_2).
\end{multline}

To simplify the following work, we define a number of integral operators. First, forward and backward integral operators,
\begin{equation}
    \hat{T}_{\perp/\|}^+[f](u)=\int_{-a}^udu_1\int_{\mathbb{S}}\,du_2\,\Xi_{\perp/\|}(u_1,u_2)f(u_2),
\end{equation}
\begin{equation}
    \hat{T}_{\perp/\|}^-[f](u)=\int^{a}_{u}du_1\int_{\mathbb{S}}\,du_2\,\Xi_{\perp/\|}(u_1,u_2)f(u_2),
\end{equation}
and in addition, a difference operator,
\begin{equation}\label{eq:82}
    \Delta \hat{T}_{\perp/\|}[f](u)=\hat{T}_{\perp/\|}^+[f](u)-\hat{T}_{\perp/\|}^-[f](u).
\end{equation}
Together, these definitions simplify Eqs.~\eqref{eq:53} and \eqref{eq:54} to
\begin{equation}\label{eq:42}
    E_{\|}(u)=E_{\|}^{\mathrm{av}}-2\pi i q\Delta\hat{T}_\perp[D_\perp](u),
\end{equation}
\begin{equation}\label{eq:43}
    D_\perp(u)=D_\perp^{\mathrm{av}}-2\pi i q\Delta\hat{T}_\|[E_{\|}](u),
\end{equation}
where $E_{\|}^{\mathrm{av}}=(E_{\|}(a)+E_{\|}(-a))/2$ and $D_\perp^{\mathrm{av}}=(D_\perp(a)+D_\perp(-a))/2$. Next, we simply insert Eq.~\eqref{eq:42} into Eq.~\eqref{eq:43} and separately Eq.~\eqref{eq:43} into Eq.~\eqref{eq:42} to cast each expression in terms of only either $E_{\|}$ or $D_\perp$,
\begin{equation}\label{eq:24}
    E_{\|}(u)=E_{\|}^{\mathrm{av}}-2\pi i q D_\perp^{\mathrm{av}}\Delta \hat{T}_{\perp}[1](u)-(2\pi q)^2\Delta \hat{T}_{\perp}\Delta \hat{T}_{\|}[E_{\|}](u),
\end{equation}
\begin{equation}\label{eq:25}
    D_\perp(u)=D_\perp^{\mathrm{av}}-2\pi i q E_{\|}^{\mathrm{av}}\Delta \hat{T}_{\|}[1](u)-(2\pi q)^2\Delta \hat{T}_{\|}\Delta \hat{T}_{\perp}[D_\perp](u).
\end{equation}
The above two equations are used to derive Eqs.~\eqref{eq:29} and \eqref{eq:40} in the main text.

\section{ETCs for susceptibilities not doubly even}\label{sc:B}
Here, we will consider the case where the susceptibilities are not doubly even. Eqs.~\eqref{eq:24} and \eqref{eq:25} are first formally inverted to isolate the field quantities,
\begin{equation}\label{eq:27}
    E_{\|}(u)=\left(1+(2\pi q)^2\Delta \hat{T}_{\perp}\Delta \hat{T}_{\|}\right)^{-1}\left[E_{\|}^{\mathrm{av}}-2\pi iq D_\perp^{\mathrm{av}}\Delta \hat{T}_{\perp}[1]\right],
\end{equation}
\begin{equation}\label{eq:28}
    D_\perp(u)=\left(1+(2\pi q)^2\Delta \hat{T}_{\|}\Delta \hat{T}_{\perp}\right)^{-1}\left[D_\perp^{\mathrm{av}}-2\pi iqE_{\|}^{\mathrm{av}}\Delta \hat{T}_{\|}[1]\right].
\end{equation}
We will define another set of auxiliary functions to simplify the coming work,
\begin{align}
    &\psi_\|(u)=\left(1+(2\pi q)^2\Delta \hat{T}_{\perp}\Delta \hat{T}_{\|}\right)^{-1}[1](u),\\
    &\psi_\perp(u)=\left(1+(2\pi q)^2\Delta \hat{T}_{\|}\Delta \hat{T}_{\perp}\right)^{-1}[1](u),\label{eq:55}
\end{align}
where the functions $\psi_{\perp/\|}$ are defined by the Fredholm integral equations,
\begin{equation}
    \psi_{\perp/\|}(u)+(2\pi q)^2\Delta \hat{T}_{\|/\perp}\Delta \hat{T}_{\perp/\|}[\psi_{\perp/\|}](u)=1.
\end{equation}

With these definitions, Eqs.~\eqref{eq:27} and \eqref{eq:28} can be written as,
\begin{equation}\label{eq:97}
    E_{\|}(u)=E_{\|}^{\mathrm{av}}\psi_\|(u)-2\pi iqD_\perp^{\mathrm{av}}\Delta \hat{T}_{\perp}[\psi_\perp](u),
\end{equation}
\begin{equation}\label{eq:98}
    D_\perp(u)=D_\perp^{\mathrm{av}}\psi_\perp(u)-2\pi iqE_{\|}^{\mathrm{av}}\Delta \hat{T}_{\|}[\psi_\|](u).
\end{equation}
These can now be directly solved for the boundary conditions by evaluating at the endpoints. However, Eqs.~\eqref{eq:97} and \eqref{eq:98} are not sufficient to fix all four endpoint values simultaneously. Take Eq.~\eqref{eq:97} for example; if we evaluate at the endpoints, we get the system
\begin{equation}\label{eq:99}
    \left[\begin{array}{ccc}
        1-\frac{1}{2}\psi_\|^+ & -\frac{1}{2}\psi_\|^+ \\
        -\frac{1}{2}\psi_\|^- & 1-\frac{1}{2}\psi_\|^-
    \end{array}\right]\left(\begin{array}{c}
         E_\|^+  \\
          E_\|^-
    \end{array}\right)=-2\pi iq D_\perp^\text{av}\left(\begin{array}{c}
         \Delta \hat{T}_{\perp}[\psi_\perp](a)  \\
          \Delta \hat{T}_{\perp}[\psi_\perp](-a)
    \end{array}\right)
\end{equation}
where $\psi_\|^{\pm}=\psi_\|(\pm a)$. The matrix in this expression is actually singular. This can be seen explicitly if we look closer at the difference operator defined in Eq.~\eqref{eq:82}. This can be simplified to a single integral operator over a sign function,
\begin{equation}\label{eq:15}
    \Delta \hat{T}_{\perp/\|}[f](u)=\int_{\mathbb{S}}du_1\int_{\mathbb{S}}du_2\,\Xi_{\perp/\|}(u_1,u_2)\sign(u-u_1)f(u_2),
\end{equation}
where splitting the bounds using the sign function recovers the difference between the forward and backward operators. This form has the added benefit of removing the variables from the integral bounds. In Eq.~\eqref{eq:15}, if $u=a$ then $u\geq u_1\,\,\,\forall u_1\in\mathbb{S}$ and if $u=-a$ then $u\leq u_1\,\,\,\forall u_1\in\mathbb{S}$ , while $u=u_1$ only on the boundary.  This allows us to evaluate the sign functions, resulting in $\Delta \hat{T}_{\perp/\|}[f](-a)=-\Delta \hat{T}_{\perp/\|}[f](a)$. This endpoint relationship suggests the definition of a shared quantity
\begin{equation}
    \Delta \hat{T}_{\perp/\|}[\psi_{\perp/\|}](\pm a)=\pm\int_{\mathbb{S}}du_1\int_{\mathbb{S}}du_2\,\Xi_{\perp/\|}(u_1,u_2)\psi_{\perp/\|}(u_2)=\pm\Xi^\mathrm{eff}_{\perp\perp/\|\|},
\end{equation}
which we shall refer to as an effective susceptibility. Consider also the definition of $\psi$ at the endpoints using the same sign function logic,
\begin{equation}
    \psi_{\perp/\|}^\pm\pm(2\pi q)^2\int_{\mathbb{S}}du_1\int_{\mathbb{S}}du_2\,\Xi_{\|/\perp}(u_1,u_2)\int_{\mathbb{S}}du_3\sign(u_2-u_3)\int_{\mathbb{S}}du_4\,\Xi_{\perp/\|}(u_3,u_4)\psi_{\perp/\|}(u_4)=1.
\end{equation}
We may once again rename the quantity common to both endpoints to find another effective susceptibility,
\begin{equation}
    \psi_{\perp/\|}^\pm=1\pm 2\pi iq\Xi^\mathrm{eff}_{\|\perp/\perp\|},
\end{equation}

where $\Xi^\mathrm{eff}_{\perp\|/\|\perp}$ is defined as
\begin{equation}
    \Xi^\mathrm{eff}_{\perp\|/\|\perp}=-2\pi iq\int_{\mathbb{S}}du_1\int_{\mathbb{S}}du_2\,\Xi_{\perp/\|}(u_1,u_2)\int_{\mathbb{S}}du_3\sign(u_2-u_3)\int_{\mathbb{S}}du_4\,\Xi_{\|/\perp}(u_3,u_4)\psi_{\|/\perp}(u_4).
\end{equation}
The reason for this particular choice of definition will become clear subsequently. This result also naturally suggests that $\psi_{\perp/\|}^++ \psi_{\perp/\|}^-=2$.

Inserting these results into Eq.~\eqref{eq:99} we get
\begin{equation}
    \left[\begin{array}{ccc}
        \frac{1}{2}(1-2\pi iq\Xi^\mathrm{eff}_{\perp\|}) & -\frac{1}{2}(1+2\pi iq\Xi^\mathrm{eff}_{\perp\|}) \\
        -\frac{1}{2}(1-2\pi iq\Xi^\mathrm{eff}_{\perp\|}) & \frac{1}{2}(1+2\pi iq\Xi^\mathrm{eff}_{\perp\|})
    \end{array}\right]\left(\begin{array}{c}
         E_\|^+  \\
          E_\|^-
    \end{array}\right)=-2\pi iq D_\perp^\text{av}\left(\begin{array}{c}
        \Xi^\mathrm{eff}_{\perp\perp}  \\
          -\Xi^\mathrm{eff}_{\perp\perp}
    \end{array}\right),
\end{equation}
which is row-dependent. This system therefore reduces to a single independent equation,
\begin{equation}
    \frac{1}{2}(1-2\pi iq\Xi^\mathrm{eff}_{\perp\|})E_\|^+-\frac{1}{2}(1+2\pi iq\Xi^\mathrm{eff}_{\perp\|})E_\|^-=-2\pi iq D_\perp^\text{av}\Xi^\mathrm{eff}_{\perp\perp}.
\end{equation}
This can then be rearranged to find
\begin{equation}
    E^+_\|-E^-_\|=-4\pi i q \left(\Xi^\mathrm{eff}_{\perp\perp}D_\perp^\mathrm{av}+\Xi^\mathrm{eff}_{\perp\|}E_\|^\mathrm{av}\right),
\end{equation}
which by symmetry tells us that the boundary condition in the other direction must be
\begin{equation}
    D^+_\perp-D^-_\perp=-4\pi i q \left(\Xi^\mathrm{eff}_{\|\|}E_\|^\mathrm{av}+\Xi^\mathrm{eff}_{\|\perp} D_\perp^\mathrm{av}\right).
\end{equation}

\bibliographystyle{unsrtnat}
\bibliography{references} 

\end{document}